\documentclass[sigconf,nonacm,10pt]{acmart}

\usepackage{natbib}
\usepackage{xspace}
\usepackage{listings}
\usepackage[draft]{minted}
\usepackage[font=small,labelfont=bf,skip=5pt]{caption}
\usepackage{subcaption}

\makeatletter
\def\maxwidth{\ifdim\Gin@nat@width>\linewidth\linewidth\else\Gin@nat@width\fi}
\def\maxheight{\ifdim\Gin@nat@height>\textheight\textheight\else\Gin@nat@height\fi}
\makeatother
\setkeys{Gin}{width=\maxwidth,height=\maxheight,keepaspectratio}
\newcommand{\parab}[1]{\noindent{\bf #1}}
\newcommand{\https}{\textit{HTTPS}\xspace}
\newcommand{\lol}[1]{}
\newcommand\nel{NEL\xspace}
\newcommand\NEL{\nel}
\newcommand\nelheader{\texttt{\nel}\xspace}
\newcommand\reportheader{\texttt{Report-To}\xspace}
\newcommand\ldns{LDNS\xspace}

\newcommand\testbed{testbed\xspace}

\newcommand\domain{neltest.com\xspace}
\newcommand\approach{our approach\xspace}
\newcommand\Approach{Our approach\xspace}

\title{Client-side Active Measurements Without Application Control}
\author{Matt Calder \quad Microsoft }
\date{}

\begin{document}
\maketitle

\section*{Abstract}
\label{sec:abstract}
Monitoring performance and availability are critical to operating successful content provider networks. Internet measurements provide data needed for traffic engineering, alerting, and network diagnostics. While there are significant benefits to combining server-side passive measurements with end-user active measurements, these capabilities are limited to a small number of content providers with both network and application control. In this work, we present a solution to a long-standing problem for a method to issue active measurements from clients without application control. Our approach uses features of the W3C Network Error Logging specification that allow a CDN to induce a browser connection to an HTTPS server of the CDN's choosing. 

\section{Introduction}
\label{sec:intro}

Web applications are ubiquitous on today's Internet -- offering quick access to popular services such as email, video-conferencing, chat, streaming media, and cloud storage. Competition is fierce amongst online cloud services and operators know that poor performance negatively impacts revenue~\cite{linden-amazon, yahoo-slow}. To ensure good user experience, content providers and content delivery networks (CDN) monitor the health of their networks and services with measurements that reflect network quality and user experience.


CDNs strive to deliver content quickly and reliably for many different web properties, but in most cases don't control the application embedding the content. 
While CDNs use a variety of measurement techniques (\S\ref{sec:other-approaches}), previous work has shown that there are significant advantages to combining
server-side and client-side active measurements.
The Odin~\cite{calder2018odin} measurement platform demonstrated the ability to construct client-LDNS mappings for high performance DNS-based redirection (\S\ref{sec:need}), true availability measurements by using fault-tolerant Internet paths, and safe network experiments from actual user populations without risk to production workloads, all of which are not possible with just server-side measurements (\S\ref{sec:other-approaches}) but are hugely valuable for performance, availability, and operations. Currently these capabilities are unique to a minority of content providers which operate a CDN and have application control (e.g. Facebook, Google, Microsoft). However, CDNs not tied to any particular application (e.g. Akamai, Fastly, Cloudflare) serve a huge amount of Internet traffic.

A general method for performing client-side active measurements without application control has been a long-standing open problem for the Internet measurement and content delivery community. In this work, we present a solution using Network Error Logging (NEL), a W3C standard for in-browser client-side measurement\cite{nelspec} and implemented in the Chromium browser. \NEL is by design a passive measurement collector providing a standard way to obtain detailed reports about performance and failures of web requests not previously possible with solutions such as the \textit{Resource Timing API}~\cite{resourcetimingapi}. A server configures \nel by setting HTTP response headers that include a \textit{report endpoint} where measurement results are uploaded. Our technique capitalizes on this report endpoint; the very action of the browser sending an \https request to a server-specified URL provides the basis for our approach.

This technique can work today on an estimated 60-69\% of web users (\S\ref{sec:nel}), enabling server-controlled measurements from previously inaccessible vantage points. Unlike existing approaches, it requires no custom JavaScript injected into web applications; the browser operates in a background process which reduces risk of web application interference. While this work is particularly relevant for CDNs, \approach improves on a general subset of web-based client-side active measurements by providing a safe and unobtrusive alternative that reduces liability of Internet researchers and operators. To the best of our knowledge, this is a novel technique and this is the first work for explore its utility. Next we discuss background on CDNs, \NEL, and existing measurement approaches.

\section{Motivation and Background}
\label{sec:motivation}



\subsection{Operational Need for Measurement}
\label{sec:need}

CDNs use Internet measurements to track performance and reliability of their services. These measurements are critically important to traffic engineering and alerting systems that react to congestion and outages on the Internet. In this work, we discuss three important scenarios:

\parab{User Performance.} 
Points-of-presence (PoP) serve content from the edge of a CDN's network, close to users, to reduce costs and improve performance. As Internet conditions change, the best performing PoP for a user may change. With these insights, automation or operator actions can then improve user experience.

\parab{Client-\ldns Association.} \textit{Redirection} is how a CDN directs users to access content from a specific PoP; usually one that provides good performance. A widely-deployed technique is for the CDN's authoritative DNS to return a PoP-specific record based on user performance, geography, or system load. In most cases, the CDN can only make a decision based on a user's local DNS (\ldns /recursive resolver) and \textit{not the user itself} so knowing the set of users served by an \ldns is critical for directing users to a good performing PoP.

\parab{Availability \& Outages.} Content providers must quickly diagnose and mitigate network and service availability issues to prevent loss of revenue from advertising, service level agreement (SLA) violations, or customers moving to a more reliable competitor.

In the next two sections, we review \nel and other approaches used understand these scenarios.

\subsection{Network Error Logging}
\label{sec:nel}

Network Error Logging (\nel) is a W3C specification\cite{nelspec} for reporting client-side networking errors and performance in web applications. \nel enables a site owner to install a measurement policy for a domain under its control by setting the \nelheader and \reportheader headers in the HTTP response as shown in Listing~\ref{listing:nelheaders}. In this example, the \nelheader header specifies a policy to sample 25\% of successful requests and 100\% of failed requests for up to 24 hours. \nel provides low-layer networking errors for failed requests that are not exposed to other JavaScript APIs. For example, \nel reports \texttt{dns.unreachable} when a DNS server cannot be reached and \texttt{tcp.timed\_out} when the TCP connection to a server times out~\cite{nelspec}.

When installed, a \nel policy applies to all resources from the same domain, and optionally sub-domain, for the \texttt{max\_age} duration. The \reportheader header's \texttt{endpoints} field is a list of \textit{report endpoints} where the browser will upload reports in a background process after a short delay, outside of the browser's critical rendering path. \NEL supports uploads using multiple redundant failover paths. If the first report endpoint is unreachable, the browser retries on subsequent endpoints~\cite{burnett2020network}.

\begin{listing}[t]
\begin{minted}{python}
NEL: {"report_to": "default", "max_age": 43200, 
      "success_fraction": 0.25, "failure_fraction": 1.0}
     
Report-To: {"group": "default", "max_age": 86400, 
     "endpoints": [{"url": "https://neltest.com/report"}]}
\end{minted}
\caption{Example \nel and Report-To response headers to configure a measurement policy.}
\label{listing:nelheaders}
\end{listing}

\NEL's first public release was in September 2018 in Chrome 69.
Microsoft Edge (79+) and other browsers based on Chromium inherit \nel functionality. \NEL covers the majority of web users due to its large browser market share as shown in Table~\ref{table:market}.


\begin{table}[t]
\footnotesize
\begin{tabular}{p{1.2cm}|p{1.0cm}|l|p{1.5cm}}
\textbf{Device Type} & \textbf{Share \%} & \textbf{Metric} & \textbf{Source}                        \\ \hline
Desktop              & 69.1             & Page Views      & NetMarketShare                         \\
Mobile               & 65.6             & Page Views      & NetMarketShare                         \\
Desktop              & 67.1             & Unique Visitors & StatCounter                           \\
Mobile               & 60.3             & Unique Visitors & StatCounter                           \\
All                  & 47.6             & Page Views      & U.S. DAP \\
All                  & 80.7             & Page Views      & w3schools.com 
\end{tabular}
\caption{Google Chrome estimated market share in April, 2020 from NetMarketShare\cite{netmarketshare}, StatCounter\cite{statcounter}, U.S. DAP\cite{usdap}, and W3CSchools\cite{w3cschools}.}
\label{table:market}
\end{table}

\subsection{Existing Approaches}
\label{sec:other-approaches}

Passive server-side measurements through web logs or packet captures are widely used by CDNs to measure performance~\cite{schlinker2019internet, schlinker2017engineering, whyhigh, latlong, chen2015end} or availability~\cite{richter2018advancing}. 
Our work also uses server-side logging to capture end-user performance but instead of organic traffic, we log traffic from \NEL report uploads.

CDNs can run active layer 3 measurements (traceroute, ping) from serving-infrastructure to users or LDNSes~\cite{zhang2010optimizing, de2017verfploeter}. These approaches can suffer from low response rates~\cite{huang2011public} and failure to test end-to-end functionality that reflects user experience such as valid HTTPS certificates. Web requests from applications can do this easily.

Web measurements from end-users can be run with JavaScript~\cite{calder2018odin,callejo2017opportunities,ahmed2017peering} or by simply embedding a transparent 1x1 pixel image in HTML. Akamai~\cite{boomerang,akamai-mpulse} and Fastly~\cite{fastly-insights} operate JavaScript measurement platforms where customers opt-in to embed a script to their application.\footnote{Injecting JavaScript is still application control, even if it is a small function within a larger application.} From speaking with several CDN operators, customer adoption is challenging. Injection of third-party code can raise performance, security, and privacy concerns~\cite{wang2013demystifying,ludin2017measuring}. Unlike JavaScript, \approach reduces interference with web application execution because it runs in a background browser process. \Approach is complementary to systems like AdTag~\cite{callejo2017opportunities}, which uses targeted ads to acquire end-user vantage points, by replacing JavaScript with less invasive measurements. \Approach also avoids issues with browser extensions preventing JavaScript ad-blockers and interruptions from users navigating away from a page.

Conviva runs measurements from end-users to different CDNs optimize video delivery for publishers~\cite{jiang2016cfa, ganjam2015c3}.

Akamai also collects measurements from its NetSession~\cite{zhao2013peer,chen2015end} platform and Media Analytics Plugin~\cite{krishnan2013video}. The Odin measurement platform is deployed in Microsoft's web and desktop applications~\cite{calder2018odin}.
These applications are installed on end-user devices, providing access to low-layer networking APIs enabling measurements such as traceroute. Coverage is subject to customer willingness to install desktop applications whereas \approach will work with any CDN customer end-user using a compatible browser.

\subsection{Benefits of Application Control}

Content providers that own end-user applications have a distinct advantage over others in their measurement capabilities.

\parab{Target Diversity.} Applications can make measurements to HTTP servers other than those serving a production user request. This can be accomplished on the web with JavaScript or embedding PoP-specific resources in HTML. This can measure the performance of alternate CDN PoPs for traffic engineering~\cite{calder2018odin, cedexis} without the need to direct a percent of users to random PoPs using DNS, potentially degrading performance. This capability is particularly desirable to avoid network experimentation with valuable production enterprise and cloud services traffic.
In cases where an alternate PoP serves users due to an outage, continuous measurements to the degraded PoP can show when the issue has been mitigated and it is safe to shift traffic back.



\parab{DNS Resolution.} Web applications don't have access direct access to DNS resolution in a browser's sandboxed JavaScript environment. This is a serious challenge considering the importance DNS plays in CDN PoP selection~\cite{chen2015end}(\S\ref{sec:need}). However, web applications can generate unique hostnames that are tied to a network property such as a client IP address. This, along with authoritative DNS control and logging, is required for client-\ldns association techniques on the web~\cite{mao2002precise, calder2015analyzing, huang2011public}.

\parab{Representative Coverage.} Application control enables coverage of the actual user population and so reflects real customer performance. Measurement platforms~\cite{ripe-atlas, thousandeyes, cedexis} with existing vantage points are helpful for monitoring but are inherently biased as they cannot represent the CDN's actual population and traffic.




These advantages are important for CDNs because they enable the critical scenarios discussed in Section~\ref{sec:need}. Next, we describe our approach which enables these measurement capabilities without application control.


\section{Approach}
\label{sec:approach}
Using the \nel HTTP header, a CDN specifies a list of endpoints where the browser will upload reports and what sampling should be applied for failed and successful requests. The main insights into our approach are that \textbf{(1)} \nel allows a server to specify an \textit{arbitrary} HTTPS URL to which the client will connect. 
\textbf{(2)} Despite its acronym, \nel can be configured to send reports for successful requests, not only failures.
\textbf{(3)} Existing server-side logging support in web servers can capture client performance~\cite{log-variables,apache}.

\subsection{Overview}
\label{sec:overview}

When a CDN serves a request to a particular user, it can strategically configure a \textit{report upload endpoint} to be a server it wants measurements for, with the user as a vantage point. The \NEL policy can specify \texttt{success\_fraction: 1.0}, which guarantees a report upload; barring the user close the browser, shut off the computer, or completely lose Internet connectivity. Reports are still delivered if a tab is closed. Server-side logging then captures user performance data from the user's report upload. In short, we induce \nel to perform an active measurement to a particular server through the report upload mechanism.

\begin{figure}[t]
    \centering
    \includegraphics[width=0.9\columnwidth]{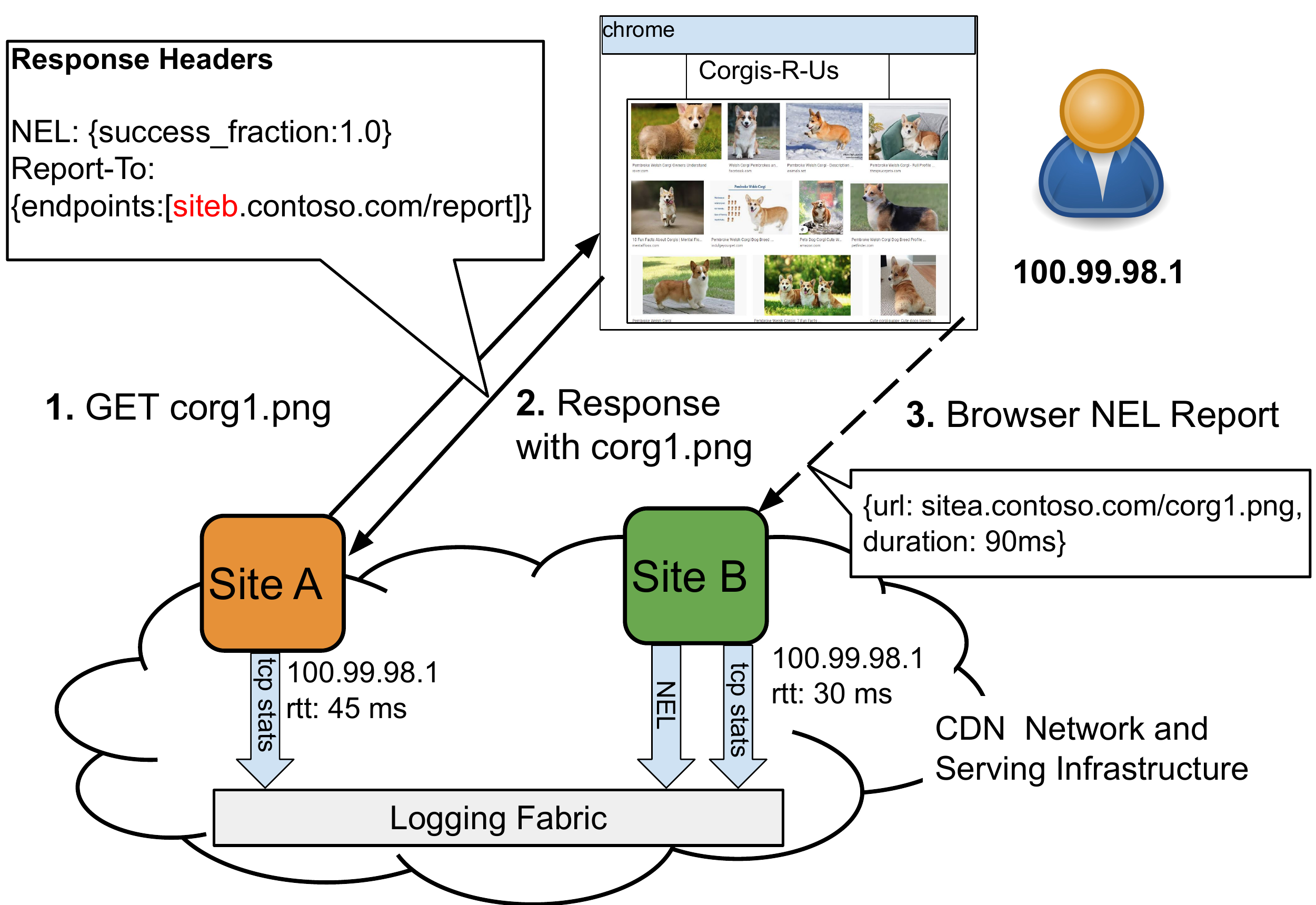}
    \caption{The CDN acquires a measurement from the user to Site B, even though Site A actively serves the user. Site A sets the Report-To endpoint to be a Site B URL. In the background, the browser uploads a \nel report to the Site B URL.}
    \label{fig:nel-alternate-pops}
\end{figure}

Figure~\ref{fig:nel-alternate-pops} illustrates the approach. 
A user with IP address \texttt{100.99.98.1} visits a popular website for hosting images. A CDN serves some of the images, which has 2 PoPs for simplicity, \textit{Site A} and \textit{Site B}. Both sites have server-side logging enabled that captures information like client IP address and TCP RTT. 

When the webpage loads, the browser resolves the hostname of the \texttt{corg1.png} image, which the CDN directs to \textit{Site A} and the browser makes a \textit{HTTP GET} request\textbf{(1)}. In \textit{Site A}'s HTTP response \textbf{(2)}, it includes the \nel header specifying that successful requests should be sampled at 100\% and that reports should be uploaded to \texttt{https://\textcolor{red}{siteb}.\domain/report}. Shortly after (within 60 seconds as of Chrome 81.0.4044.129) the image is received, 
the browser makes a \textit{HTTP POST} with the \NEL report data to \texttt{https://\textcolor{red}{siteb}.\domain/report}.

The browser controls the \NEL report upload timing and at the time of this writing there is no server or user settings to modify the upload interval or start time.  

\subsection{NEL Policy Considerations}
\label{sec:nel-policy}

There can be only one NEL policy per-domain, per-browser installation. If a subsequent resource from the same domain is loaded later with a different NEL policy, the previous policy is overwritten. Any resource timing data queued for upload will be sent to the endpoint in the most recent policy for the domain, irrespective of the policy present at the time the resource was loaded. Chrome uploads reports on a 60-second timer which means that for a single browser, it is possible to receive one report upload per domain every minute. The report payload may contain multiple \nel resource timing measurements depending on the sampling configured in the policy.

\subsection{Report Endpoint Sampling}

A naive server-side strategy for configuring \reportheader, such as setting a random report endpoint, may result in unpredictable report endpoint sampling.
One strategy may be to choose a report endpoint based on a hash of the current minute and client IP address (or user id) modulo the number of report endpoints. This way, the measurement endpoint is predictable and with a large user population, there will be high measurement volume per endpoint in aggregate over an ISP, geography, or prefix. For more frequent measurements per user, multiple sub-domains may be used support multiple \NEL policies at the same time. $N$ sub-domains enables $N$ report endpoint samples per minute. 

\subsection{Limitations}

While \approach offers new measurement capabilities from many web vantage points, it has some limitations we describe here.

\parab{Delay.} NEL reports are uploaded within 60 seconds of a resource fetch. This delay may not be acceptable for all applications.

\parab{Targets.} We use \nel to trigger an HTTP request to a particular server via the report endpoint but performance must still be captured on the server-side, so measurement targets are limited to those under the experimenter's control.

\parab{Metrics.} Currently \approach is well suited for capturing latency which is critical for a large number of web applications. We leave extensions for other important metrics such as jitter, packet loss, and throughput to future work.

\parab{Overhead.} \nel sends additional traffic by including additional response headers and bandwidth for uploading reports. This can be mitigated by server-side header sampling that considers network and device type when generating \nel policies.

\parab{Web-Only.} \Approach is currently limited to browsers implementing the \nel specification. 

\subsection{Privacy}
\NEL was designed with privacy in mind and \approach does not subvert the following principals from Burnett et al.~\cite{burnett2020network}: (1) A server cannot collect information about end-users it does not already have access to. (2) No measurements are performed outside of regular user activity (3) End users can opt-out of \NEL. (4) Only site operators can configure collection for their site and where reports are uploaded. 
\section{Applications}
\label{sec:applications}

In this section we describe applications of \approach.

\subsection{Alternate PoP Measurements}

Directing users to a nearby PoP for good performance is fundamental to the CDN business~\cite{nygren2010akamai}(\S\ref{sec:need}). Because network conditions on the Internet change often and quickly, CDNs use network performance data to decide which PoP or path will provide a user with the best performance~\cite{chen2015end, yap2017taking, schlinker2017engineering, calder2015analyzing}.

Figure~\ref{fig:nel-alternate-pops} demonstrates how \approach can support alternate PoP measurements, albeit with a two site example. This extends easily to a server in any PoP setting the report endpoints to different PoPs in the CDN network. Since CDNs can have 100s or 1000s of PoPs, a control system is needed to program response headers for useful, coordinated measurements based on client network and geographic characteristics. \Approach removes the need to do network experiments to alternate PoPs with production traffic, reducing harm to end-user performance and allowing measurement for cloud and enterprise traffic populations.


User to PoP measurements provide fine-grain per-client IP address performance data but the unit of redirection control for many CDNs is the user's LDNS (\S\ref{sec:need}) so next we look at how \approach can measure client-LDNS association.

\subsection{Client-LDNS Association}
\label{sec:app-client-ldns}

\begin{figure}[t]
    \centering
    \includegraphics{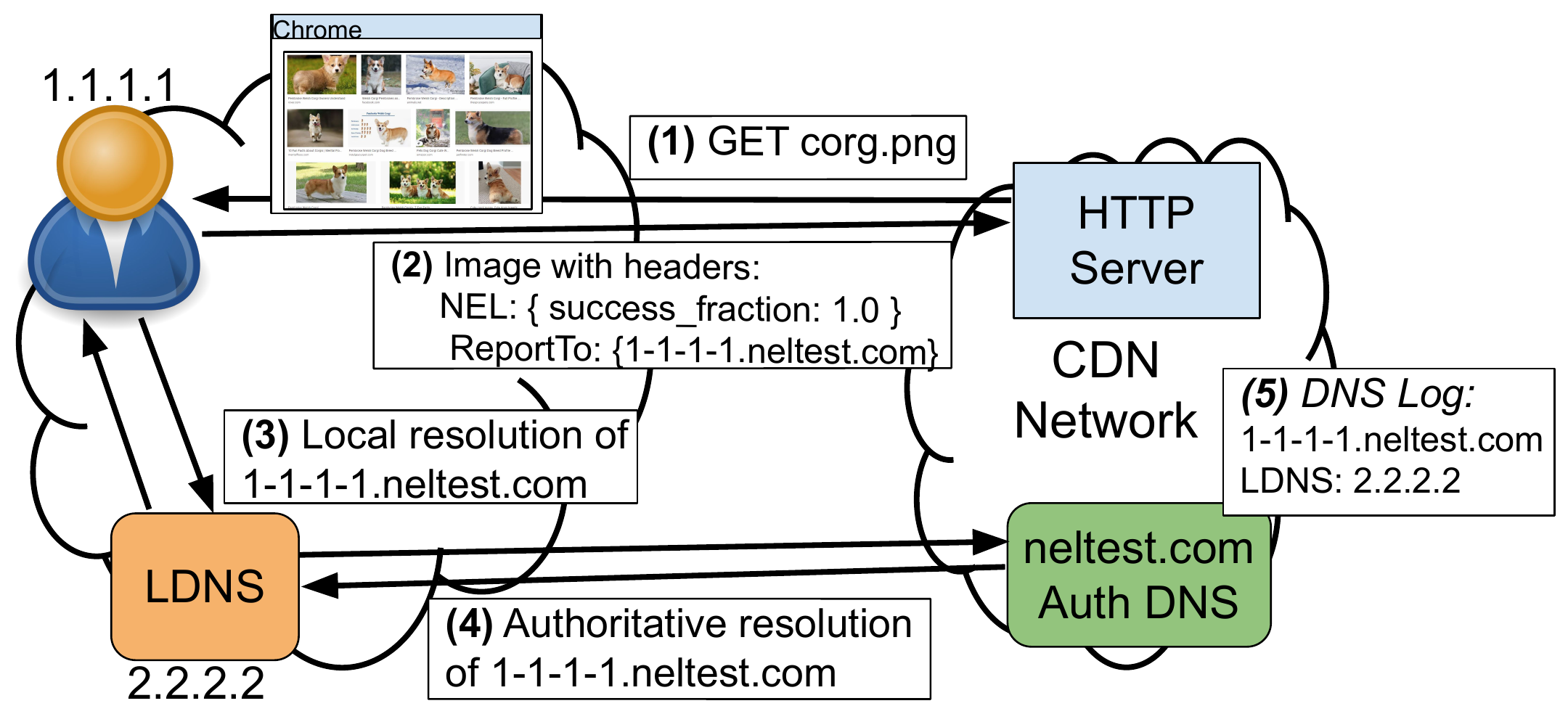}
    \caption{A CDN can associate client and LDNS IP addresses by generating custom domains server-side, and forcing the client browser to resolve it for the \NEL report upload.}
    \label{fig:nel-client-ldns}
\end{figure}

Understanding the relationship between clients and their LDNSes is also critical for good CDN performance(\S\ref{sec:need}). Figure~\ref{fig:nel-client-ldns} shows how to capture this with \approach. A user with IP address \texttt{1.1.1.1} is viewing an image sharing website. The browser fetches one of the images \textbf{(1)} from a CDN server. The server responds with the image data, along with \textbf{(2)} a NEL policy and the Report-To endpoint set to \texttt{https://1-1-1-1.neltest.com}, with the user's IP address encoded in the CDN's hostname~\cite{mao2002precise}, which the server knows from the request's remote socket IP address. When the browser starts the NEL report upload process \textbf{(3)}, it will resolve \texttt{https://1-1-1-1.neltest.com}, which will first be forwarded to the client's LDNS (2.2.2.2), and assuming no cached record exists, will forward the request to the CDN's authoritative resolver \textbf{(4)}. The authoritative DNS logs capture the association through the client's IP address in the hostname to resolve, and the LDNS's IP address from which it received the request \textbf{(5)}.


\subsection{Availability Monitoring}

A CDN must monitor PoP availability to know if client traffic is unable to reach it. Monitoring can be detected by drops in traffic volume~\cite{richter2018advancing}, client-side active measurement~\cite{calder2018odin}, or commercial monitoring platforms~\cite{cedexis, thousandeyes}. When an availability issue is detected, CDNs can mitigate the issue by moving traffic to an alternate PoP. But without additional probes to verify a fix, users cannot be safely switched back without the risk of blackholing traffic. By continuing to measure the degraded PoP with actual user web requests, the CDN can verify that end-to-end connectivity works before shifting traffic back. If continuous measurement was already in place, one could verify that the number of successful uploads to the degraded PoP is returning to historically ``normal'' values. 



\section{Evaluation}
\label{sec:evaluation}

In this section we demonstrate the feasibility of \approach with a proof-of-concept \testbed.

\subsection{Testbed Setup}

Our \testbed consists of four virtual machines (VM) serving as HTTP servers deployed in four different Microsoft Azure regions~\cite{azure-regions}, with two of the VMs also serving as authoritative DNS for our test domain. The VMs run in the West US2 (Seattle), North Central US (Chicago), France Central (Paris), and Southeast Asia (Singapore) regions that were chosen to provide diverse geographic locations and cost-effectiveness for budget constraints. DNS runs in West US2 and France Central.

The HTTP servers run Ubuntu 18.04 with NGINX 1.14.0 and PHP support. We configured NGINX access logs to record client IP address, requested URL, POST body, request duration, RTT\cite{log-variables}. 
A simple PHP script acts as a controller for dynamically setting NEL response headers when the HTTP server receives any request that is not a NEL report upload. We run a custom Python DNS server that logs the LDNS IP and request to a Redis database.

We performed all tests on an Ubuntu 18.04 laptop using Chrome version 81.0.4044.129 from a home broadband network in Pacific Northwest (PNW) region of the United States. Because \approach is dependent on a browser with \nel support, we cannot make use of existing non-browser measurement platforms such as RIPE Atlas~\cite{ripe-atlas}. We instead use the Private Internet Access (PIA) VPN service to evaluate measurements from geographically diverse vantage points~\cite{pia}. We plan to make all code and configuration available on GitHub.


\subsection{Alternate Region Measurements}
\label{sec:eval-alternate-regions}

We first look at how \approach captures changes in user performance to multiple cloud regions. To test this, we use a simple webpage that embeds four images. Each image URL has the scheme \texttt{<nel-region>.westus2.\domain}, which always serves the image content from West US2 but signals to our controller to set the Report-To endpoint to be the \texttt{<nel-region>}. This allows us to receive a measurement per region every minute (\S\ref{sec:overview}).
The page is reloaded every 60 seconds. Our experiment begins with no VPN connection, then we switch to Chicago, Paris, and Singapore VPN gateways, then turn off VPN again.

\begin{figure}[t]
    \centering
    \includegraphics{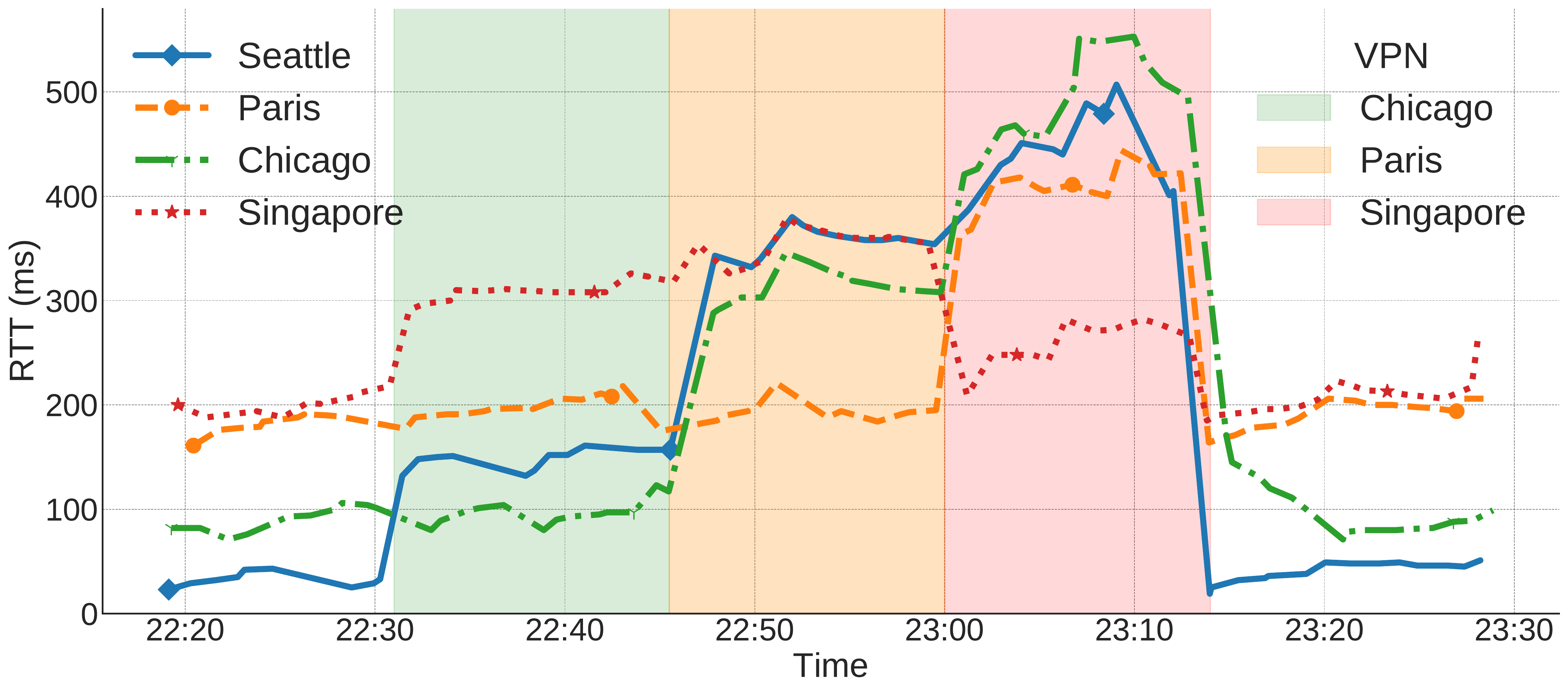}
    \vspace{-15pt}
    \caption{Performance from a single vantage point to four cloud regions with VPN gateways changed every 15 minutes.}
    \label{fig:nel-timeseries}
\end{figure}

Figure~\ref{fig:nel-timeseries} shows the performance observed from the PNW home broadband network to the four cloud regions in our \testbed over ~70 minutes with four different VPN configurations. We extract regional RTT data directly from each region's NGINX access logs. In the first 10 minutes, we see that the lowest latency region is Seattle, followed by Chicago, Paris, and then Singapore, which is expected given our PNW vantage point. At approximately \textit{22:30}, we enabled the VPN for the Chicago gateway and see Seattle and Singapore RTTs spike, and Chicago becomes the lowest latency region. Chicago and Paris latency remains similar since the VPN tunnel travels in the direction as the non-VPN path. In contrast, the path to the Seattle and Singapore regions have greatly inflated RTTs since they must make a trip East to Chicago before traveling back West. We observe similar patterns with the Paris and Singapore VPN gateways. When the VPN is turned off at the end of the experiment, per-region RTTs return to similar range as the experiment start.

This experiment validates that \approach can be used to monitor end-user performance with browser client-side measurements and that client-side network changes are quickly and accurately reflected in server-side logs.

\subsection{\NEL Client Responsiveness}

\begin{figure}
    \centering
    \includegraphics{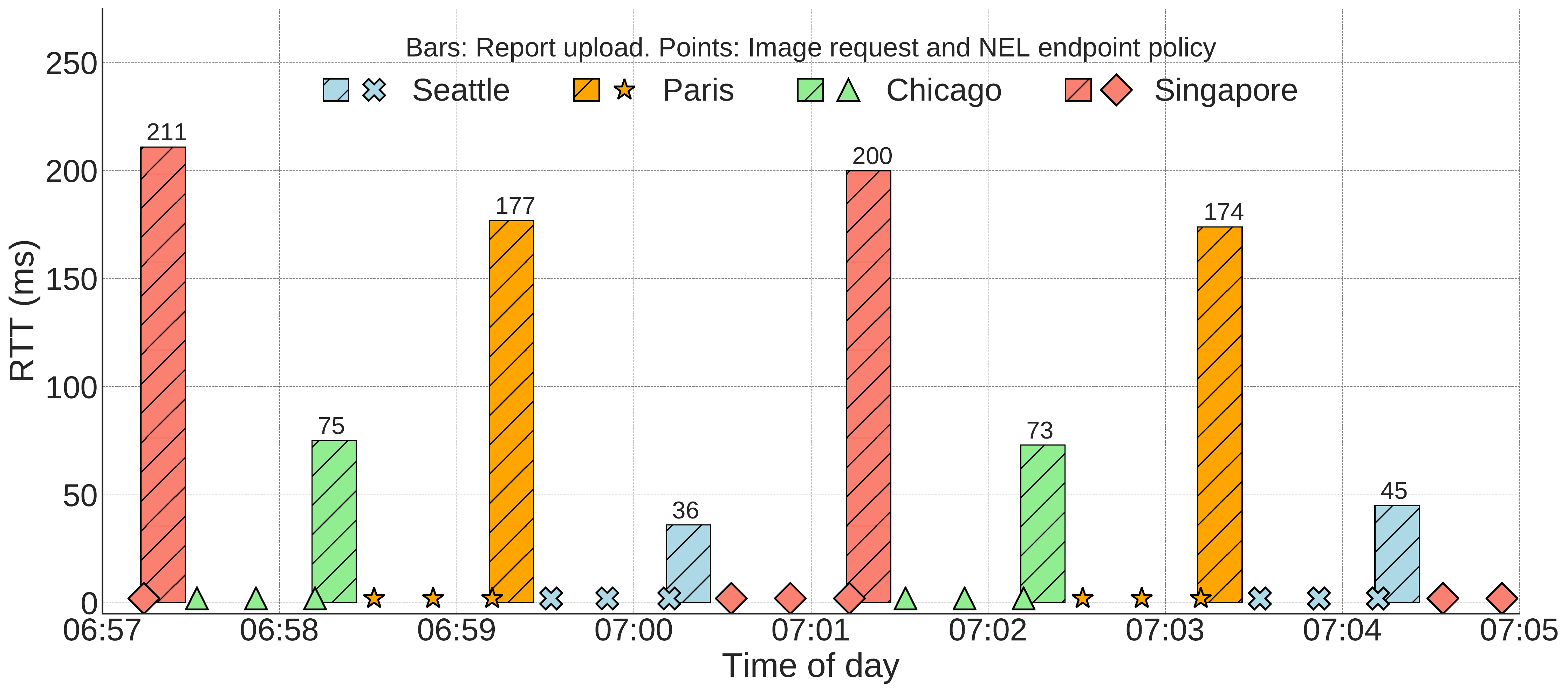}
    \caption{Chrome's response to \nel policy changes over time. Report uploads are always sent to the region from the most recent policy.}
    \label{fig:nelpolicy}
\end{figure}

Next, we examine a CDN's ability to (re)specify \NEL policies on a client. We configured our test webpage to reload an image from the same domain (single \nel policy) every 20 seconds and the server-side controller responds with a different report endpoints every minute but a fixed \texttt{max\_age} of 300 seconds. Figure~\ref{fig:nelpolicy} shows the report endpoint measurement results over an 8-minute window, showing 2 full cycles for each upload region. The points at $0$ along the x-axis show the report endpoint region sent from the server with each image request. We can see that within the same minute, all 3 images contain the same report endpoint. At the end of the browser's 1 minute report interval, it sends reports to the region from the \textit{latest} \nel policy as shown by the bars and their corresponding RTTs. This demonstrates that when a policy is updated, the new one overwrites existing policies for the same domain, even if the previous policy's \texttt{max\_age} has not expired.

\vspace{-10pt}

\subsection{Client-LDNS Association}

\begin{figure}
    \centering
    \includegraphics[]{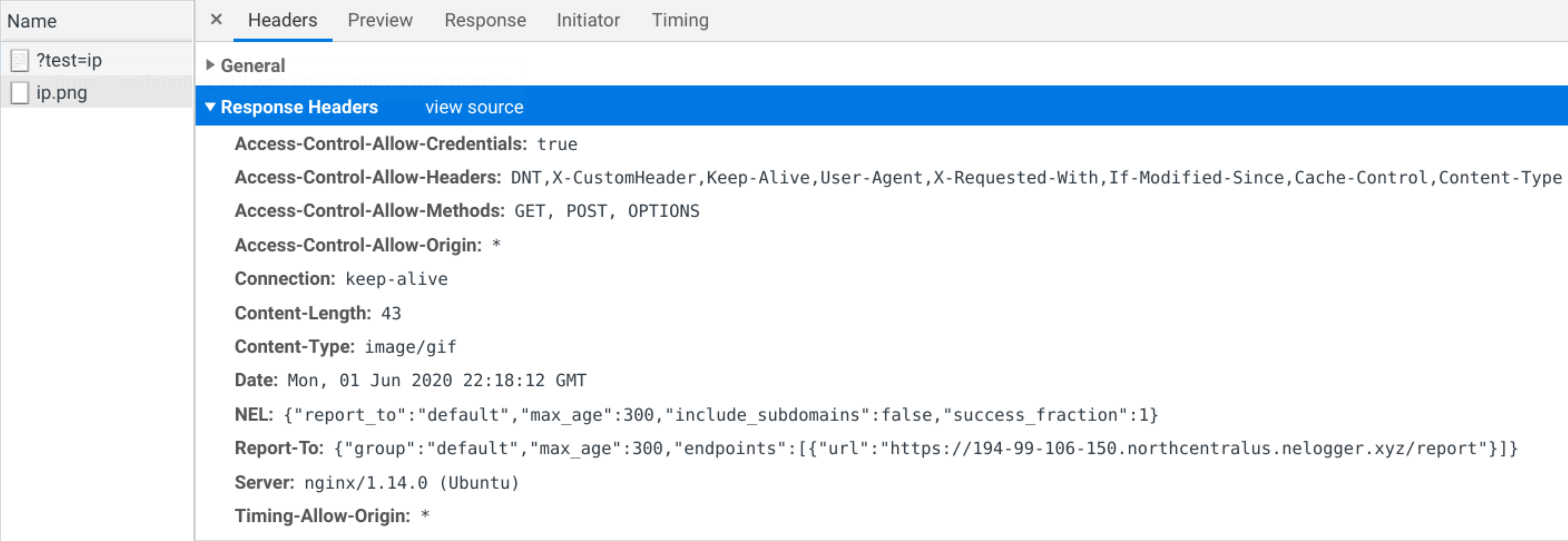}
    \caption{Chrome developer tools showing image response headers when connected to the France VPN gateway. The VPN's IP address is encoded into the Report-To endpoint.}
    \label{fig:chrome-dev}
\end{figure}

\begin{figure}
    \centering
    \includegraphics[width=0.65\columnwidth]{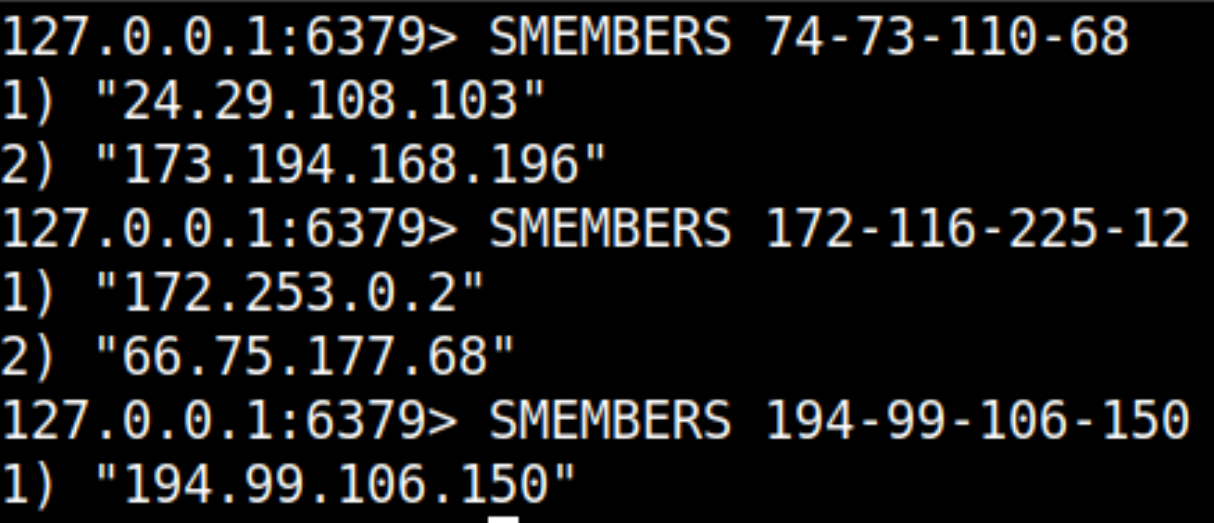}
    \caption{Redis cli showing the LDNS IP addresses for 3 client IP address keys.}
    \label{fig:redis}
\end{figure}

In this section, we show a proof-of-concept for capturing client-LDNS association. We used the approach described Section~\ref{sec:app-client-ldns} and requested two volunteers on the east and west coast of the U.S. to load a test page once with their ISP's DNS configured and another with Google Public DNS. The browser response view and DNS log capture is shown in Figures~\ref{fig:chrome-dev} and \ref{fig:redis} and Table~\ref{table:client-ldns} summarizes the data. Coincidentally, both volunteers are on Spectrum broadband but we can see their LDNSes are very different. As expected, we capture Spectrum LDNSes talking to our authoritative resolvers but they are in completely different /8 prefixes. The West US client's IP and LDNS are still associated with a Time Warner Cable ASN from a recent acquisition. When the clients configured Google Public DNS, we see that the East US client was directed to the Washington D.C. instance while the West US client was sent to Los Angeles~\cite{google-public-dns}. Lastly, if we look at the VPN client in France, we see that the client IP and LDNS are the same, which is what we want to see, as DNS leakage is an issue with some VPNs~\cite{khan2018empirical}.


\begin{table}
\footnotesize
\setlength{\tabcolsep}{2pt}
\renewcommand{\arraystretch}{0.85}
\begin{tabular}{p{2.0cm}|p{1.35cm}|p{1.9cm}|p{2.85cm}}
\textbf{Client} & \textbf{Client ISP} & \textbf{LDNS} & \textbf{Description}   \\ \hline \hline
74.73.110.0/24 (East US) & Spectrum   & 25.29.108.103 & Spectrum LDNS \\ \hline
 -                       & -          & 173.194.168.196 & Google PDNS in Wash. D.C. \\ \hline
172.116.225.0/24 (West US) & Spectrum & 66.75.177.68 & TWC (Spectrum) LDNS \\ \hline
-                          & -        & 172.253.0.2  & Google PDNS in L.A. \\ \hline
194.99.106.150 (VPN France) & M247    & 194.99.106.150 & M247 LDNS
\end{tabular}
\caption{Three examples of client-LDNS association made by our prototype. This data reflects the Redis database in Figure~\ref{fig:redis}.}
\label{table:client-ldns}
\end{table}


\section{Conclusion}
\label{sec:conclusion}
In this work we describe the first approach that enables a CDN, cloud, or content providers to initiate client-side measurements to their serving infrastructure without application control. We describe how to use \approach to perform several measurements critical to CDN operations such as alternate PoP measurements, client to LDNS association, and availability monitoring. Our cloud-based testbed and evaluation demonstrates \approach working in practice.

\bibliographystyle{ACM-Reference-Format.bst}
\bibliography{bib.bib}

\end{document}